\documentclass{elsart}

\usepackage{psfig}

\usepackage{natbib}

\begin{document}

\runauthor{Cicero, Caesar and Vergil}


\begin{frontmatter}

\title{Studies of the high luminosity quasar, PDS 456}
\author{J.Reeves$^{1}$, P.O'Brien$^{1}$, S.Vaughan$^{1}$, 
D.Law-Green$^{1}$, M. Ward$^{1}$} 
\author{C. Simpson$^{2}$, K. Pounds$^{1}$, R. Edelson$^{1,3}$}
\address{1) X-Ray Astronomy Group; Department of Physics and Astronomy;
Leicester University; Leicester LE1 7RH; U.K. \\
2) Subaru Telescope, National Astronomical Observatory of Japan, 650 N.
A`oh\={o}k\={u} Place, Hilo, HI 96720, U.S.A. \\
3) Department of Physics and Astronomy; University of California,
Los Angeles; Los Angeles, CA 90095-1562; U.S.A. \\}

\begin{abstract}

X-ray and multi-wavelength observations of the most luminous known
local (z~$<0.3$) AGN, the recently discovered radio-quiet quasar
PDS~456, are presented. The spectral energy distribution shows that
PDS~456 has a bolometric luminosity of 10$^{47}$~erg/s, peaking in the
UV. The X-ray spectrum obtained by {\it ASCA} and {\it RXTE} shows
considerable complexity. The most striking feature observed is a deep,
highly-ionised, iron K edge (8.7 keV, rest-frame), originating via
reprocessing from highly ionised material, possibly the inner
accretion disk.  PDS~456 was found to be remarkably variable for its
luminosity; in one flare the X-ray flux doubled in just
$\sim15$~ksec. If confirmed this would be an unprecedented event in a
high-luminosity source, with a light-crossing time corresponding to
$\sim2R_{S}$.  The implications are that either flaring occurs within
the very central regions, or else that PDS~456 is a `super-Eddington'
or relativistically beamed system. 

\end{abstract}


\end{frontmatter}

\section{Introduction}

PDS 456 is a bright, {\it radio-quiet} QSO (V=14) recently discovered by
Torres {\it et al.} (1997), in a search for young stellar objects. 
It lies fairly close to the Galactic plane
($\beta=12$) and is seen through an extinction of A$_{V}=1.5$. PDS 456
is at a similar redshift (z=0.184) to 3C 273, but has a higher
bolometric luminosity (by a factor of $\times$1.7, Simpson et al. 1999).  
Overall, PDS 456 is the most luminous
object in the local (z~$<0.3$) Universe 
(with M$_{V}=-27$, L$_{BOL}\sim10^{47}$~erg~s$^{-1}$). 

\begin{figure}
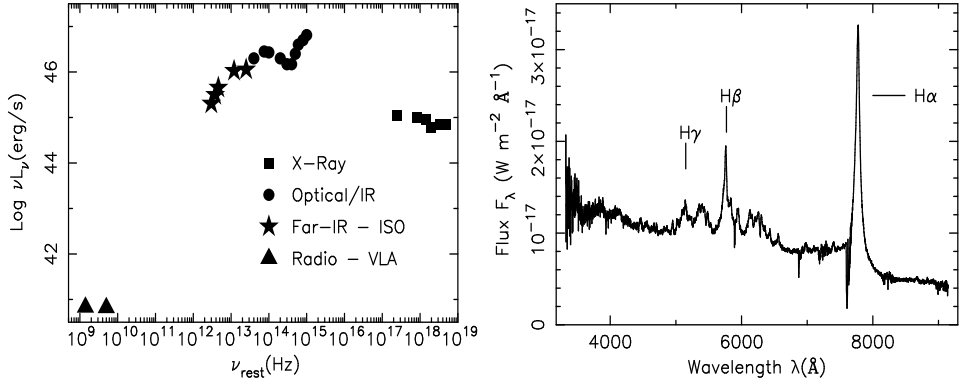

\centerline{
\hbox{
\psfig
{figure=reeves_fig1a.ps,width=0.45\textwidth,height=0.22\textheight,angle=270}
\psfig
{figure=reeves_fig1b.ps,width=0.45\textwidth,height=0.22\textheight,angle=270} 
}}

\caption{(a) The radio to X-ray SED of PDS~456; the emission peaks in the
blue/UV part of the spectrum. It is seen that the bolometric
luminosity of PDS~456 approaches 10$^{47}$~erg~s$^{-1}$. (b) The optical
spectrum of PDS 456.}

\end{figure}

\section{Multi-Wavelength Properties of PDS~456}

We have conducted an extensive campaign to observe PDS 456 from the
radio through to the hard X-ray band. The spectral energy distribution
(SED) is shown in Figure 1a and the optical spectrum in Figure 1b.  
PDS 456 shows strong H I emission lines, 
and like many NLS1s has strong optical Fe\textsc{II}
emission but weak O\textsc{III}. Although by definition a
broad-line quasar, PDS 456 has only a moderate width in H$\beta$ (FWHM = 3000
km s$^{-1}$).  
VLA observations confirm that, unlike 3C 273, PDS 456 is radio-quiet
(R$_{L}$=-0.7) and has little extended radio emission. 
Overall the SED is dominated by the optical/UV `big blue bump'. 
The bolometric luminosity of PDS~456 is of the order
10$^{47}$~erg~s$^{-1}$ (assuming 
$ H_0 = 50 $~km\,s$^{-1}$\,Mpc$^{-1}$ and $ q_0 = 0.5 $).

\vspace{-0.5cm}
\section{X-ray Observations of PDS~456}

\subsection{The X-ray Spectrum}

As part of our campaign, we observed PDS~456 with ASCA on 7-8 March
1998 and with RXTE on 7-10 March 1998. 
The hard X-ray spectrum of PDS 456 obtained by both ASCA and RXTE
shows complex features (see Reeves {\it et al.} 2000 for details). The
data/model residuals from a power-law fit ($\Gamma=2.4$) to the RXTE
data are shown in Figure 2a. 
Unsurprisingly a power-law gave an inadequate fit to the data in this band.
We find that an unusually deep and ionised Fe K edge is observed, with 
best-fit parameters of E=8.7$\pm$0.2 keV and $\tau=0.75\pm0.15$. 
The edge is detected in both the ASCA and RXTE data to $>$99.99\% confidence.
There is also some evidence in the X-ray spectrum for a broadened 
($\sigma\sim1$~keV) line
at ~6 keV; this line may originate from the inner disk, as 
hypothesised in Seyfert 1s (Tanaka {\it et al.} 1995). A `warm'
absorber of lower ionisation may also be present at soft X-ray
energies, whose properties are also similar to those observed in 
Seyfert 1s (e.g. Reynolds 1997). 

A model consisting of reflection off a highly ionised accretion disk 
provides the best-fit to the hard X-ray spectrum; with disk solid angle,
R=$\Omega/2\pi=1.0$, ionisation parameter, $\xi=6000$~erg~cm~s$^{-1}$ and
T$_{disk}=10^{6}$~K.  The high-ionisation of the disk reflection 
component can reproduce both the depth of the edge and its energy at 8.7
keV.  Such high-ionisation reflection features are predicted in disk
photoionisation models (e.g. Ross, Fabian \& Young 1999, Nayakshin et
al. 1999), particularly at high accretion rates when the primary
X-ray emission is steep. 
Therefore the high ionisation of the reflector could imply a high
accretion rate in PDS~456 (relative to the Eddington limit), particularly as
$\xi\propto\dot{m}^{3}$ for a photoionised accretion disk.  
This interpretation is consistent with the
other X-ray properties of PDS~456, namely a steep underlying X-ray continuum
and rapid X-ray variability, both of which are commonplace in NLS1s
(Boller {\it et al.} 1996). NLS1s are also thought to
be accreting near the Eddington limit (e.g. Pounds et al.
1995); indeed recent evidence has been found in one NLS1 (Ark 564) for
a spectrum consistent with ionised disk reflection (Vaughan {\it et al.}
1999).  

\begin{figure}
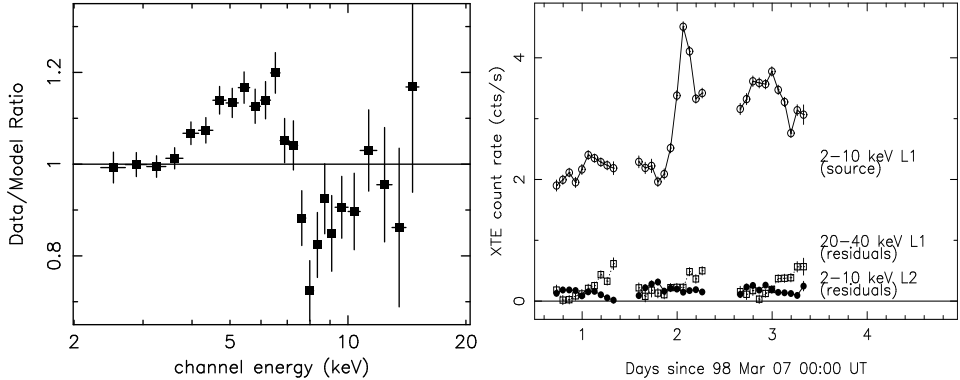

\centerline{
\hbox{
\psfig
{figure=reeves_fig2a.ps,width=0.45\textwidth,height=0.22\textheight,angle=270}
\psfig
{figure=reeves_fig2b.ps,width=0.45\textwidth,height=0.22\textheight,angle=270}
}}

\caption{(a) The data/model ratio residuals from a simple
($\Gamma=2.4$) power-law fit to the {\it RXTE} data of PDS 456. The effect
of the deep, ionised iron K edge is clearly seen in the residuals.
(b) The 2-10 keV {\it RXTE} lightcurve of PDS~456. The source appears
to flare by a factor of 2.1 in 17 ksec, corresponding to a
light-crossing size of $\sim2R_{S}$ under the assumptions of isotropy
inherent in the Eddington equation.} 

\end{figure}

\subsection{X-ray Variability}

Both the ASCA and RXTE data were examined to search for X-ray
variability.  A strong hard X-ray flare is observed in the RXTE
observation (figure 2b), well above any residual fluctuations in the
detector background; the doubling time for the flare was
$\sim15$~ksec. (Note that, unfortunately, the shorter ASCA observation 
had ended by the time of the flare.) 
We also calculated that the probability of finding another
contaminating source of comparable brightness in the RXTE beam was low
($<2$\%), although not totally excluded. Additionally there is no
other X-ray source detected in the RASS (ROSAT All Sky Survey) to
within a degree of PDS 456.  

Therefore, if confirmed, this would be unprecedented behaviour in
such a high-luminosity source.  This suggests, from simple
light-crossing arguments, a maximum size of $l=4.5\times10^{12}$~m for
the varying region. For a black hole of mass $10^{9}$M$_{\odot}$
(corresponding to PDS~456, with L$_{BOL}=10^{47}$ erg~s$^{-1}$, at the
Eddington limit), this implies that the X-ray flare occurs {\bf within a
region of less than 2 Schwarzschild radii} (2R$_{S}$).  
A smaller mass black
hole would loosen this requirement somewhat, but would then imply a
super-Eddington accretion rate. {\it Therefore one possible
implication of the rapid variability is accretion near to or greater
than L$_{Edd}$.} The variability also implies a (non-beamed)
efficiency of converting matter to energy of $\sim$~5\%, close to the
limit for a Schwarzschild black hole (see Fabian 1979).

\section{Conclusions}

In conclusion, PDS~456 is a remarkable object, showing clear features
of a high ionisation reprocessor, one possible interpretation of which
is through reflection off a highly ionised accretion disk. Overall the
high ionisation spectral features, steep X-ray emission and the
extreme rapid variability suggest that the super-massive black hole in
PDS~456 could be running at an unusually high accretion rate.

\end{document}